\documentclass[prd,preprint]{revtex4}
\usepackage{graphicx}
\usepackage{hyperref}
\usepackage{amsmath}
\usepackage{times}

\begin{document} 

\title{Simulating central forces in the classroom}
\author{
Juan M.\ Aguirregabiria 
} 
\email{juanmari.aguirregabiria@ehu.eus}
\affiliation{Theoretical Physics, 
University of the Basque Country (UPV/EHU), \\
P.~O.~Box 644,
48080 Bilbao, Spain
}

\bigskip

\begin{abstract}
We describe the easy to customize and extend open source Java program \emph{Central force workbench}
that can be used in the classroom 
to simulate the motion of a particle (or a two-body system) under central
forces. It may be useful to illustrate problems in which the analytical solution
is not available as well as to help students to grasp a more intuitive
understanding of the main features of this kind of problem and to realize
the exceptional nature of the particular case of Newtonian (and harmonic) forces.
We also include some suggestions on how to use the program as a pedagogical tool.
\end{abstract} 


\maketitle

\section{Introduction}\label{sec:intro}

When physics students are introduced to central forces  the subject is
mainly, if not exclusively, focused on Newtonian forces. While this is completely
justified by the physical importance (and historical influence) of Newtonian forces,
it is not always stressed the exceptional nature of this kind of force law among
all the possible central forces. I have often found students that think any
orbit of positive mechanical energy is hyperbolic, even if the force is not
Newtonian. Of course, this belief can be dispelled by telling them that in
a harmonic potential $V\propto r^2$ every orbit is elliptic and has positive
energy. However, this in turn can reinforce a seemingly natural tendency to think
that bounded orbits are always elliptic or, at least, closed, while it is a
consequence of a dynamical symmetry 
\cite{bb:symmetry,bb:mtnez,bb:goldstein3}
the closely related \cite{bb:grant} Newtonian and harmonic force laws.

Even in the Newtonian case there is room for misunderstandings. When dealing with
the two-body problem one is lead to separate the (trivial) motion of the
center-of-mass and the relative motion of the two bodies.
After studying the Keplerian orbits, many students are perplexed if
asked
how would look the orbit of each body in the center-of-mass system
or, worse,
in another inertial frame in which the total linear momentum is not null
(as we should expect to  see in the motion of a binary star system).
This is clearly an instance in which a simulation may be helpful.

On the other hand, force laws for which there is a closed analytical solution for
the orbit equation
are exceptional (although of paramount importance): apart from the Newtonian and
harmonic cases, one can mention the potential $V=-k/r+a/r^2\ ${\cite{bb:goldstein4}}
and the special
relativistic version of the Kepler problem, \cite{bb:boyer} which can be solved by
using
essentially the same computation needed for Newtonian forces. 
The Cotes' spirals \cite{bb:Whittaker} are even more remarkable because not only the orbit equation
but also the time evolution may be solved by elementary methods.
There are other
cases in which the orbit equation can be solved
in terms of special functions, \cite{bb:goldstein} but they are of little use in
introductory courses.
Finally there is a handful of cases in which the solution (or, at least, the orbit
equation)
can be computed in terms
of elementary functions for special values of the initial conditions 
(often when the  mechanical energy vanishes or equals the maximum value of the
effective potential).

For these reasons a versatile computational tool that can be easily used in
the classroom (and by students at home) to simulate the motion under central 
forces can be a valuable pedagogical asset. The aim of this work is to
present such a tool, which we will succinctly  do in Section~\ref{sec:program}, while
Section~\ref{sec:activities} will be devoted to suggest how to effectively use
it as a pedagogical tool in a introductory course on mechanics.
The Appendix provides a complete reference for users of the program.

\section{The program}\label{sec:program}

\emph{Central force workbench} is written in the freeware platform \emph{Easy Java/Javascript Simulations}  
by Francisco Esquembre. \cite{bb:paco} A single file, \texttt{central.jar}, \cite{bb:jar} 
contains everything:
the executable code, the source code and the complete help system. It can be 
run as a standalone program in any computer with Java 1.5 or newer. 
The same file can also be
embedded in a web page. \cite{bb:web} The program is open source and the source code 
may be extracted from \texttt{central.jar} and easily extended and customized
by means of  \emph{Easy Java/Javascript Simulations}, \cite{bb:paco} which can also be used to
add translations to other languages (the current distribution includes
English and Spanish).  

\begin{figure}
\begin{center}
\includegraphics[width=\textwidth]{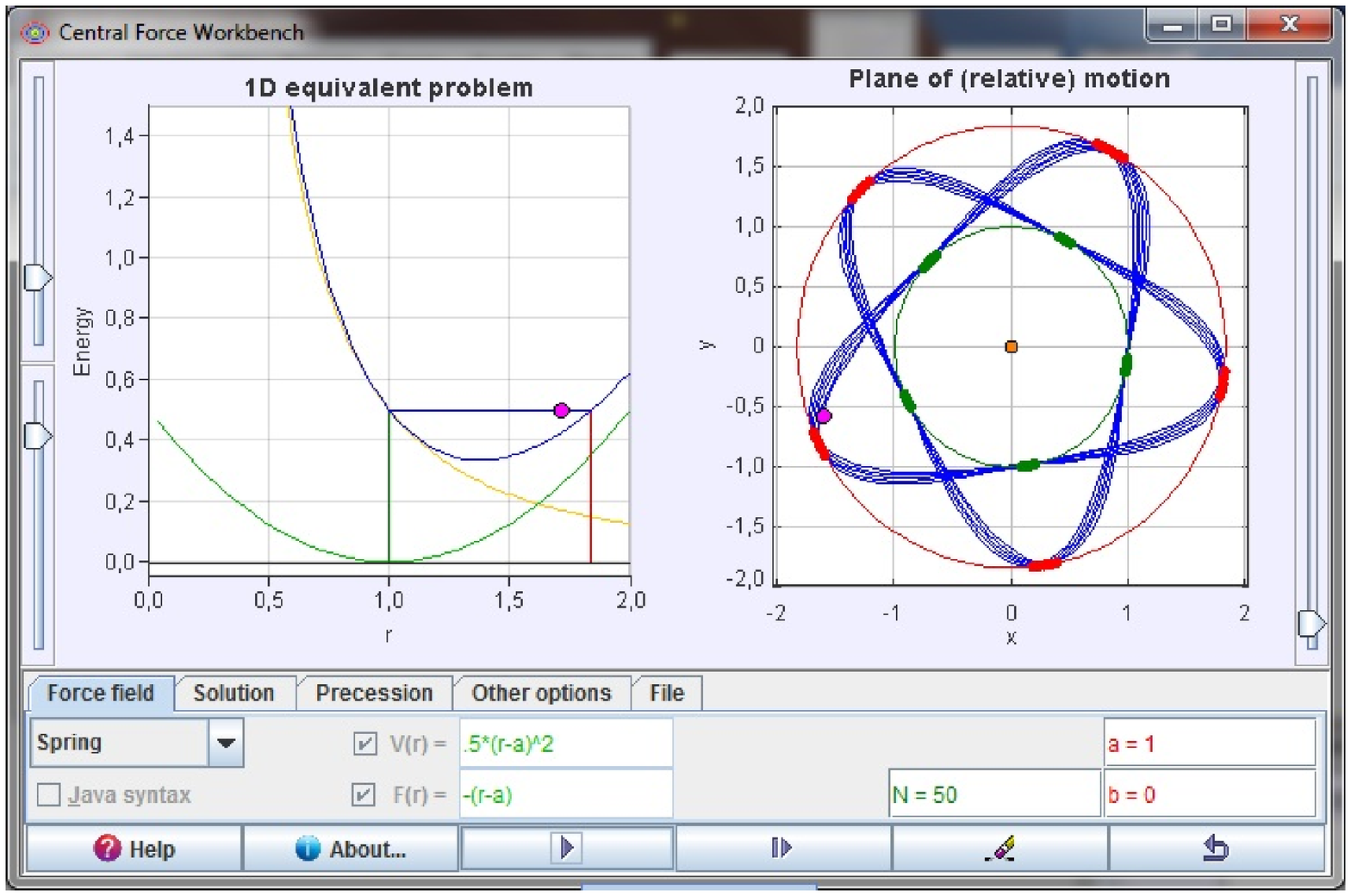}
\end{center}
\caption{The main window of the simulation.\label{fig:main}} 
\end{figure}

The main window, shown in Figure~\ref{fig:main}, displays on the left
the potential $V(r)$ in green, the centrifugal term $L^2/2mr^2$ in orange and the 
effective potential $V_\mathrm{e}(r) \equiv V(r) +L^2/2mr^2$ in blue. 
($L$ is the angular momentum and $m$ the particle mass or 
the reduced mass for the two-body system). 
There will appear the evolution in the equivalent one-dimensional problem 
corresponding to the radial motion.
The two-dimensional motion of the single particle 
(or the relative motion in the two-body problem) is displayed on the right. 

\begin{figure}
\begin{center}
\includegraphics{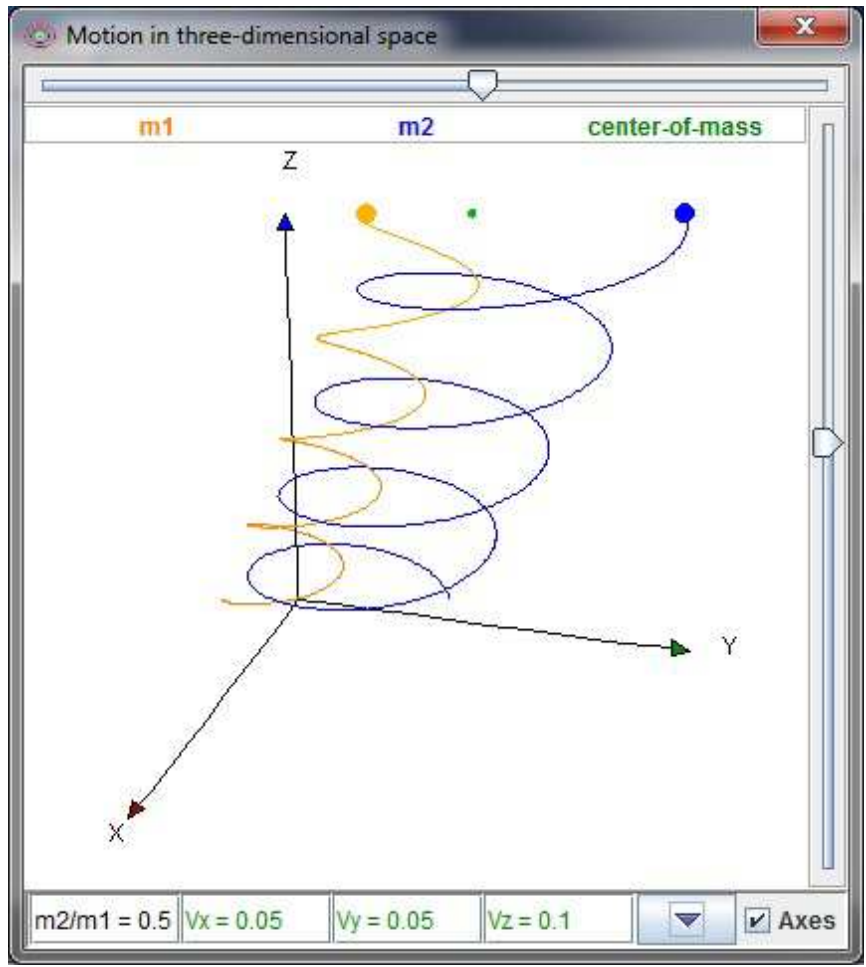}
\end{center}
\caption{Three-dimensional projection of a two-body system
evolution.\label{fig:3d}}
\end{figure}

Optionally, a projection of the three-dimensional motion of both particles in the
two-body problem can be seen in another windows (see Figure~\ref{fig:3d}).
There one can select the projection direction in either the
center-of-mass
system or in another inertial frame in which the center-of-mass moves
with prescribed constant velocity.

\begin{figure}
\begin{center}
\includegraphics[width=\textwidth]{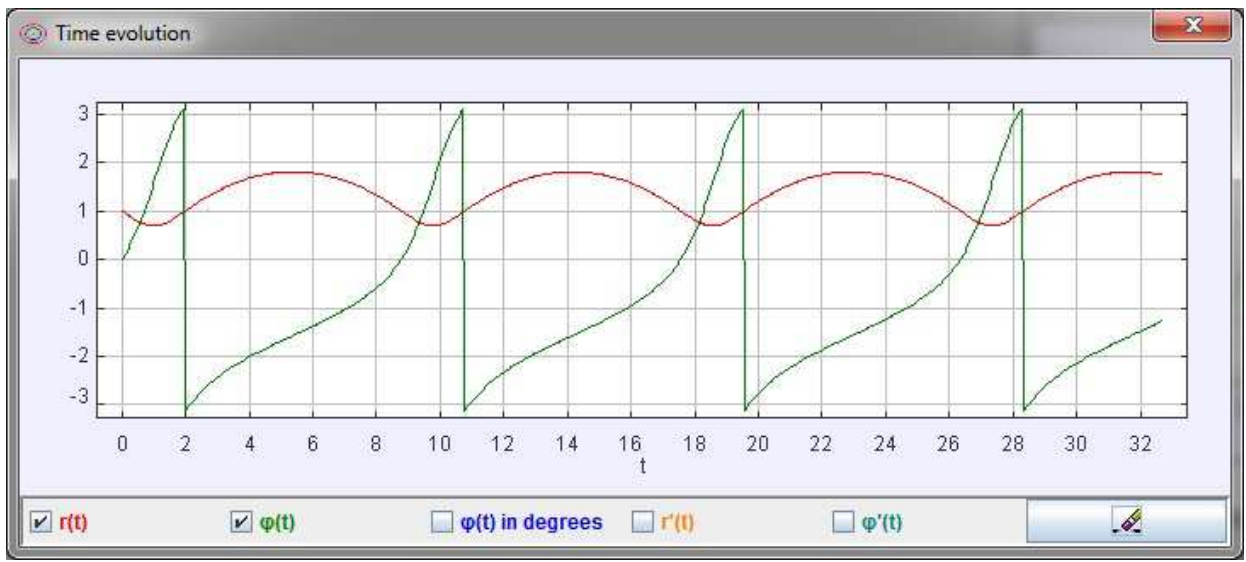}
\end{center}
\caption{Time evolution of polar coordinates.\label{fig:time}}
\end{figure}

It is also possible to display the time evolution of the polar
coordinates $r$ and $\varphi$ and their time derivatives, as shown in 
Figure~\ref{fig:time}. Furthermore, all the results 
displayed in the different windows can be captured in a video sequence or 
exported in several graphics
 formats, including Encapsulated PostScript. The solution
 points can also be exported  in a numerical
table and a tool to perform data analysis 
(including Fourier analysis) of the solution points is embedded in the program:
just right click on an orbit and choose the appropriate menu entry.

The force law can be selected from a list of predefined examples, but the user is
free to enter its own law or to modify one of the predefined ones. Physical
parameters (initial conditions, energy, angular momentum and parameters in the
force law), options for the numerical code (integration method and tolerance) and
display settings can be easily changed. It is also possible to make the program
display and compute the orbit apses, in order to compute precession, for instance. The
program is user-friendly (an informative tooltip is displayed when the mouse
hovers over an element) and includes a complete help system, with a
troubleshooting section.  

We refer the interested reader to the Appendix (or to the help
system) for more information on using the program.

\section{Suggested activities}\label{sec:activities}

The simulation provides a general framework to explore motion in
central force fields. By selecting \textbf{User defined} in the \textbf{Force field} tab
the user may define its own problem, maybe starting from one of the predefined ones.
We list below the predefined force fields, which
we have used in the classroom, in the form
of exercises to be solved by means of the program and, in some cases, by
using analytical methods.
After a short formulation of each exercise, we add a short comment.
(In the examples included in the program the units are chosen so that $k=m=GM=1$.)

\begin{description}

\item[Newtonian]:  ``Simulate the Keplerian orbits,
in the potential $V(r) = -k/r$, 
for different values of the mechanical energy $E$ and angular momentum $L$. 
Consider both the attractive
case ($k > 0$) and the repulsive force ($k < 0$).''
\\\textbf{Comment}: The program can be useful for illustration
purposes.

\item[Harmonic]: ``Check that every
orbit in the harmonic  potential $V(r)=\frac12k r^2$ is elliptic if $k>0$. Discuss the
difference between
these elliptic orbits and the Newtonian ones. Which among Kepler's laws
are
still satisfied and how do we have to change the remaining? What happens
if $k<0$?''
\\\textbf{Comment}: As mentioned before, this can be used to stress
that positive mechanical energy does not imply hyperbolic orbits. The simulation
may be also used instead of a formal proof of the elliptic nature
of the orbits.

\item[Bertrand]: ``Check
that among the potentials in the form $V(r)= k r^a$ only in the Newtonian case
$a =
-1$ and
in the harmonic case, $a = 2$, all bounded orbits are closed (i.e.,
periodic). You may try $a = -1,\ -1.02,\ -0.98,\ 2.02, \ldots$  Try finding
isolated periodic orbits
(the \textbf{Precession} tab may be useful when using trial and
error to do this).''
\\\textbf{Comment}: This can be used to introduce
precession and, especially, to discuss Bertrand's theorem.
\cite{bb:bertrand,bb:brown,bb:goldstein2}
Students should realize the exceptional nature of Newtonian and harmonic
forces, where a dynamical symmetry (the Laplace-Runge-Lenz vector or the
equivalent Hamilton eccentricity vector) makes
closed every bounded orbit.
\cite{bb:symmetry,bb:mtnez,bb:goldstein3}

\item[Spring]: ``A point mass moves on a smooth table,
attached to a spring
of proper length $a$. Show that the mass moves in the central potential $V(r) =\frac12k (r-a)^2$. 
Compute some orbits. Try finding
periodic orbits.''
\\\textbf{Comment}: Another modification of the harmonic force which gives
very different
orbits (see Figure \ref{fig:main}). This is a simple example in which the force is
both attractive and
repulsive along many orbits.

\item[Constant force]: ``A particle sliding without friction
inside a vertical conical surface, or two particles connected by a
string passing through a small hole in a horizontal smooth table).
Show that the particle moves in the central potential $V(r) =k r$,
for appropriate $k$. Try
finding periodic orbits, including circular ones.''
\\\textbf{Comment}: A couple of equivalent well known examples. \cite{bb:amalio,bb:sacks}

\item[Galaxy]: ``In a naive model of the motion of a star around a galaxy 
of radius $R$, total mass $M$ and constant density, the force
for $r \le R$ is harmonic,
$F(r)= -GMmr/R^3$, and for $r \ge R$ Newtonian, $F(r) = -GMm/r^2$.
Check that the orbits
are periodic ellipses if they are inside the radius $R$ and, in
general open if they cross that value.''
\\\textbf{Comment}: A mix of harmonic and Newtonian cases, \cite{bb:jiang} which
again shows
the exceptionality of those cases.

\item[Yukawa]:
``Discuss, in terms of the angular
momentum $L$, the existence and stability of circular orbits  in the Yukawa potential $V(r) = -k\exp(-r/a)/r$ from nuclear physics.''
\\\textbf{Comment}: An illustration of unstable circular orbits.

\item[Ion-atom]: ``Check that in the potential $V(r) = -k/r^4$ for the ion-atom interaction,
the orbits with energy $E = 0$ are circles of radius $(km/2L^2)^{1/2}$ going
through the force center. Which is the maximum impact parameter $b =
L/(2mE)^{1/2}$ for capture? Discuss the orbits with $E = L^4/16km^2$: find their
analytical expression.''
\\\textbf{Comment}: The simplest case of circular orbit with the geometric center
outside the force center (the remaining cases are discussed in \cite{bb:petsch}).
The strong singularity at $r=0$ provides an opportunity to discuss difficulties
in numerical simulations.

\item[Cotes]: ``Find the orbit equation $r(\varphi)$ and the time
evolution, $(r(t),\varphi(t))$, of a particle moving in the potential $V(r)=-k/r^2$.'' 
\\\textbf{Comment}: Everything can be calculated in terms of elementary functions.
The orbits a called Cotes' spirals. \cite{bb:Whittaker}

\item[Open orbits]:  ``Check
that orbits in the potential $V(r) = -k/r + a/r^2$, with $a > -L^2/2m$ with
negative energy are periodic (i.e., closed) when $a =
L^2(p^2/q^2-1)/2m$, for
$p,\ q = 1,2,\ldots$ These orbits have $p$ pericenters (and $p$ apocenters)
every
$q$ turns around the force center (or the other particle in two-body
problems). Use the \textbf{Precession} tab to check the analytic result for
precession: $\Delta\varphi = 2\pi [(1+2ma/L^2)^{-1/2}-1]$.'' 
\\\textbf{Comment}: This is
an interesting example because a full analytical solution for the orbit equation
is easy to obtain,
\cite{bb:goldstein4}
which gives exact values for precession and existence of closed orbits.

\item[Precession]: ``The potential $V(r) = -a/r(1+b/r^2)$, with $a =GMm$ and $b= L^2/m^2c^2$,
 describes
the
motion of a test particle in a static gravitational field with spherical
symmetry (i.e., in the Schwarzschild metric). \cite{bb:hobson1} Since for Mercury
(for
instance) $b$ is very small, the general relativistic contribution to the
perihelion precession is easier to compute by analytical (approximate)
methods. However, one can use the simulation for periastron precessions
in  strong gravitational fields (i.e., for $b$ not so small). Use the
\textbf{Precession} tab to check the analytic (approximate) result for
precession: $\Delta\varphi \approx 6\pi m^2a^2b/L^4$. Notice that in this
case
$t$ is the proper
time along the particle worldline.''
\\\textbf{Comment}: An illustration from General Relativity.

\item[Light]: 
``The potential $V(r) = -a/r^3$ , with $a = GMh^2/c^2$, describes the motion of a photon in a static gravitational 
field with spherical symmetry (the Schwarzschild
metric). \cite{bb:hobson2} Here $t$ is the
affine parameter and
$h = r^2d\varphi/dt$ a conserved quantity. The light deviation by the Sun
 is too
small: you'd better use a not so small value for $a$. Use
the \textbf{Precession} tab to check the analytic (approximate) result for light
deviation: $\Phi\approx 4GM/bc^2$, where $b$ is the impact parameter. In
our notation $m=1$,
$h =L = b(2E)^{1/2}$ and $\Phi\approx  4a(2E)^{1/2}/L^3$.''
\\\textbf{Comment}: The second example from General Relativity.

\end{description}

\section{Conclusions}\label{sec:concl}

We have presented a versatile, portable, customizable and free program that can be a
useful resource for teaching central forces. Some ideas to effectively use
the tool in the classroom have been included.

\section*{\large Appendix: Program reference}\label{sec:cref}

This appendix  has been taken from the help system
and can be skipped on first reading, but it might
be useful to fully appreciate the many features in the program.

\subsection{Description of the program}

This simulation explores the motion of a particle (or a two-body system)
in a central potential $V(r)$. The user can select the latter from a
number of built-in examples or define a new one.

\begin{itemize}
\item The unit mass is the particle mass (or the reduced mass for the
two-body problem), so that $m = 1$.

\item At the bottom of the main window there is a panel with five tabs.
In \textbf{Force field} you select the force field to study and in 
\textbf{Solution} the
initial conditions and the corresponding mechanical energy $E$ and angular
momentum $L$ as well as other options.

\item The \textbf{Precession} tab may be used show the orbit apses and to compute
the angular distance between successive pericenters and/or apocenters.

\item The fourth tab contains \textbf{Other options} described below.

\item The fifth tab, \textbf{File}, allows saving and retrieving the current
simulation state, if the program is run standalone (not included as an
applet in an HTML page).

\item On the bottom there are the buttons to play/pause the simulation,
perform a single integration step, erase the computed orbits and values
and reset the program settings.

\end{itemize}

With this simulation you may:
\begin{itemize}

\item Run it as an independent program (in any computer with Java 1.5 or
newer) or as an applet from a web page and choose its look and feel in
\textbf{Other options}.

\item Select the force field from a list of ready-to-run examples,
modify them or enter your own definition in the \textbf{Force field} tab.

\item Change easily parameters, initial conditions and output settings.

\item See, on the left, the potential $V(r)$, the centrifugal term $L^2/2mr^2$
and the effective potential $V\text{e}(r) \equiv V(r) +L^2/2mr^2$, 
as well as the
evolution in the one-dimensional equivalent problem corresponding to the
radial motion.

\item See, on the right, the two-dimensional motion of the single
particle around the fixed center of force (or the relative motion in the
two-body problem).

\item Optionally see, in a separate window, the time evolution given by
$r(t)$, $\varphi(t)$, $\dot r(t)$ and $\dot\varphi(t)$.

\item Optionally see, in a separate window, a projection of the
three-dimensional motion of both particles in the two-body problem (or
the single particle in a fixed force field, if \textbf{m2/m1} = 0) in an inertial
frame. The latter may be the center-of-mass system or different from it.

\item Optionally show the orbit apses and compute the angular distance
between two successive pericenters and/or apocenters, in order to
evaluate precession.

\item Capture graphics and video sequences.

\item If run as an independent program, get in a table and analyze
solution data, save to the disk and retrieve from there the full state
of a simulation (see below, in \textbf{Saving and restoring the simulation}), and
many more, from the menu that opens when right clicking on an orbit.

\item When a numerical parameter has to be entered, use a full
mathematical expression instead of a number. For instance, you may write
\texttt{sin(pi/4)} instead of \texttt{0.7071...}

\item Try different numerical routines to integrate the equations of motion.

\item Get a tooltip on any element by putting the mouse cursor over it.

\item It is open source and can be easily improved or customized by
using the freeware platform \emph{Easy Java/Javascript Simulations}. \cite{bb:paco} In fact the source
code is included in the distribution jar file and can be retrieved by
using the menu that opens when right clicking on the upper window.

\item Currently English (default) and Spanish versions are included, but
other languages can be easily added by using \emph{Easy Java/Javascript Simulations}. \cite{bb:paco} To
switch the language use the menu that opens when right clicking on the
upper window.

\end{itemize}

\subsection{Selecting the force field}

\begin{enumerate}

\item In the \textbf{Force field} tab you may select one of the predefined
examples or (after selecting \textbf{User defined}) modify it or enter your own
force field.

\item In the latter case, teh best and faster results are obtained when
you specify both the potential $V(r)$ and the force $F(r)$. Of course, you
must make sure $F(r) = -V '(r)$; otherwise the displayed potential graphs
will be wrong. You may use the parameters $a$ and $b$ in the definitions of
$V(r)$ and $F(r)$.

\item If you do not specify $F(r)$, the program will compute it as $-V '(r)$
by means of an algorithm using the classical central-point formula along
with Richardson's extrapolation (\emph{deferred approach to the limit}). This
will work in most cases, but since the numerical derivative is
essentially unstable, you may need to change the starting value $\Delta r$ used
by the algorithm. Sometimes a larger value will work and in other cases
a smaller value will be necessary.

\item If you do not specify $V(r)$, the program will use Romberg's method
to compute it as $V(r) = -\int F(r)\, dr + C$. The integration constant is
fixed by the user selected values for a reference point $r_0$ and the
corresponding value $V_0 = V(r_0)$. Since the numerical integration may be
time-consuming, depending on your system's performance, you may notice
slight delays in the refreshing of the potential graphs and limits when
changing settings (but not while playing the simulation which will use
the $F(r)$ you provided). You can speed up things by selecting a lower
value for the number of points ($N$) in each graph (or even setting $N = 0$ to
avoid seeing them).

\item On the left there appear (in the form of lines with $N$ points) the
potential $V(r)$, the centrifugal term $L^2/2mr^2$ and the effective potential
energy $V_\mathrm{e}(r) \equiv V(r) +L^2/2mr^2$.

\item Put the mouse pointer over an element to get the corresponding tooltip.
\end{enumerate}
\subsection{Syntax}
\begin{itemize}

\item By default, the syntax used to define $V(r)$ and/or $F(r)$ is
fairly standard in computing. For instance, to define $-0.5 r^2 |\cos r|$
you have to enter \texttt{-0.5*r\^{}2*abs(cos(r))}. If the syntax is wrong or
incomplete, the definition background will become red when \texttt{Enter} is
pressed to validate the expression.

\item Each definition may contain numbers, the radial coordinate $r$, the
user-defined parameters $a$ and $b$, arithmetic and exponentiation operators
($+$ $-$ $*$ $/$ \^{}) and elementary functions such as $\cos$, 
$\tan$, etc. The energy $E$
and the angular momentum $L$, as well as other internal variables, are
also available, but it is not advisable to use them to avoid unexpected
(and difficult to understand) results.

\item If you know some Java programming, you may prefer using its
notation. Just select \textbf{Java syntax} and remember that the previous example
may be written as \texttt{-0.5*Math.pow(r,2)*Math.abs(Math.cos(r))}. 
Probably the
only advantage of this syntax is that a few more mathematical functions
are available (see \cite{bb:java}.)

\end{itemize}

\noindent\textbf{Example}

Let us consider a naive model of the motion of a star around a galaxy of
radius $R$, total mass $M$ and constant density. The force for $r \le R$ is
harmonic, $F(r) = - GMmr/R^3$, and for $r \ge R$ Newtonian, $F(r) = -GMm/r^2$.

\begin{itemize}

\item As usual we select as unit mass the star mass $m$.

\item Our unit length will be the radius $R$.

\item The unit time will be $(R^3/GM)^{1/2}$.

\item In these units $m = GM = R = 1$. 
(Of course, we could choose other constant values for these magnitudes
in order to change the simulation scale).

\item The force may be written as $F(r) = -\min(r,r^2)$, by means of the
function $\min$ which computes the minimum of two values, and we will let
the program the task of computing the potential.

\item See the \textbf{Galaxy} predefined example.
\end{itemize}

\subsection{Computing the solution}

To choose the simulation parameters, activate the \textbf{Solution} tab.
\begin{itemize}

\item On the left there appear (in the form of lines with $N$ points) the
potential $V(r)$, the centrifugal term $L^2/2mr^2$ and the effective potential
energy $V_\textrm{e}(r) \equiv V(r) +L^2/2mr^2$. 
You may use the mouse (or the controls in
the \textbf{Solution} tab) to select the mechanical energy $E$ and the initial
polar distance $r$. The vertical range can be changed with the sliders on
the left and the horizontal range with the slider on the right. The
graphical evolution of the energy $E$ (as well as its numerical value)
provides a good measure of the quality of the numerical integration: it
should remain (nearly) constant (but it may appear a pixel off the
horizontal due to the finite screen resolution and unavoidable round-off
errors).

\item The two-dimensional motion of the single particle (or the relative
motion in the two-body problem) is displayed on the right. You may use
the mouse to select the initial position: the program will automatically
set $r$ and $\varphi$ (which also may be entered in the numerical controls) and, if
necessary, $E$. The horizontal and vertical ranges can be changed with the
slider on the right. The axes type can be changed in the entry \textbf{Axes in
plane} (in the tab \textbf{Other options}).

\item The accuracy of the integration routine is $\varepsilon$ and the step (i.e.,
the interval between solution points) $\Delta t$. Check \textbf{Limits} to see the
maximum and minimum values of $r$  and \textbf{Orbits} to draw the orbits (and not
only the position).

\item If \textbf{3D} is checked, a projection of the three-dimensional motion of
both particles in the two-body problem (or the single particle in a
fixed force field, if \textbf{m2/m1} = 0) in an inertial frame will be shown. The
center of mass moves with velocity $(V_x,V_y,V_z)$ = \textbf{(Vx,Vy,Vz)} in that
frame, which will be the center-of-mass frame if \textbf{(Vx,Vy,Vz)} = (0,0,0).
The center-of-mass is displayed in green, the lighter particle in blue
and the heavier body in orange. You may drag the mouse to change the
projection direction, the distance to the projection screen (while
holding down the \texttt{Shift} key), and the displayed range (while holding down
the \texttt{Ctrl} key). The point of view may also be changed by using the
sliders.

\item If \textbf{r(t), $\boldsymbol\varphi$(t)} is checked the time evolution is
displayed in a separate window. You may select there the functions to be
displayed among $r(t)$, $\varphi(t)$ (in radians or degrees),
 $\dot r(t)$ =\textbf{r'(t)} 
and $\dot\varphi(t)$ = \textbf{$\boldsymbol{\varphi}$'(t)}.

\item Selecting \textbf{Use L} will instruct the simulation to try using Kepler's
second law: the particle will move faster (slower) when the distance
from the force center is smaller (larger). Notice, however, that the
result may not be perfect, since the numerical routine might need much
more time for some range of distance.

\item You can change the numerical routine used to compute the solution
in the \textbf{Other options} tab. In most cases the default Cash-Karp (5) will
work fine; but you can experience using other methods. It might be
necessary to change $\Delta t$ in the \textbf{Solution} tab. 
For a description of the available
methods see \cite{bb:solvers}. 

\item Put the mouse pointer over an element to get the corresponding tooltip.

\end{itemize}

\subsection{Precession}

The tab \textbf{Precession} may be used to show the orbit apses and to
compute the angular distance between successive pericenters and/or
apocenters, in order to compute their precession. It is also useful to
find (nearly) periodic orbits by trial and error.

\begin{enumerate}

\item Check \textbf{Pericenters} to show their positions and \textbf{Values} to compute
their angular position.

\item Play the simulation.

\item When the system goes through a pericenter it will be shown as a
green point (if \textbf{Pericenters} is checked) and, if \textbf{Values} is
checked, its value will be displayed in radians and degrees.

\item When another pericenter is reached, the angular distance between
the last two ---in the range $(-\pi,\pi)$--- will be shown, in radians and
degrees. This will be the precession value (at least if there is a
single pericenter in each turn).

\item The same procedure may be used with \textbf{Apocenters} (which will be displayed in red).

\end{enumerate}

\subsection{Deflection}

To compute deflection in a scattering process (as in the case of light
rays grazing the Sun surface in the \textbf{Light} example)
proceed as follows:

\begin{enumerate}

\item Select a large value for $r$ (maybe 1000).
\item Set $\varphi = -180^\circ$.
\item Uncheck \textbf{dr/dt $>$ 0}.
\item Play the simulation.
\item As $r$ increases the angle $\varphi$ will approach (slowly) the scattering angle.
\end{enumerate}

\subsection{Saving and restoring the simulation}

When the program is run as a standalone program (but not, for security reasons, when 
it is
an applet in an HTML page) one may save to the disk the current
simulation state to retrieve it later from there.

\noindent\textbf{To save the current state to the disk}

\begin{enumerate}

\item	Use the \textbf{Save state now} button in the \textbf{File} tab
to select the destination file.

\item	You can also use the menu that opens when right clicking on the
simulation window; but in that case make sure \textbf{User defined} is selected
in the \textbf{Force field} tab. This won't change other settings, but forgetting
to do it will probably give undesired results, for reading a different
force field from the disk will read the corresponding default settings,
instead of the ones saved to the disk.

\end{enumerate}

\noindent\textbf{To restore from the disk a simulation state}

\begin{enumerate}

\item	Use the \textbf{Load state now} button in the \textbf{File} tab
to select a previously saved state file.

\item	Alternatively use the menu that opens when right clicking on the
simulation window to select a previously saved state file.

\end{enumerate}

\noindent\textbf{When the program (not the applet) exits}
\nopagebreak[4]

\begin{itemize}

\item If \textbf{Ask the user} in the \textbf{Other options} tab is checked, the user will
be given the opportunity to save the current simulation state to a file,
from where it can be retrieved later from the menu that opens when right
clicking the main window.

\item If the entry next to \textbf{Save state to} (which can be used to select a
destination file) in the \textbf{Other options} tab is not empty, the program
will automatically save there the current state, which will be
automatically read when the next session starts. Notice that the \textbf{Force
field} will be changed to \textbf{User defined} in order to make sure every other
option is restored (see above).

\end{itemize}

\subsection{Troubleshooting}

In such a versatile program, with so many settings, there is some room
for trouble.

\begin{itemize}

\item You may use the sliders on the left to select the maximum and
minimum energy values and the slider on the right to change the range of
the radial coordinate $r$. If your graphs are still out of range you can
change the length and time units with the value of $L$ and appropriate
constants in $V(r)$ and/or $F(r)$.

\item If the simulation is very slow, try increasing the integration
step length $\Delta t$. Changing the required accuracy in the numerical
routine (the value in $\varepsilon$) may also help: the quality of the integration
should be right while the blue orbit on the left is horizontal, i.e.,
while the mechanical energy remains constant (the angular momentum is
always constant).

\item If the orbits are displayed the simulation will become
increasingly slower, for it has to draw more and more points. Erasing
the orbits   from time to time, will help.

\item Depending on your system performance, you may notice the
simulation freezes while the Java runtime machine is carrying out a
garbage collection.

\item If the simulation seems to do nothing when played, you may try
changing the integration step length $\Delta t$, to give a better starting point
to the numerical method. If $F(r)$ is not specified, the problem might
also be the value for $\Delta r$.

\item You should also check your definitions for $V(r)$ and $F(r)$ to make
sure they are valid expressions (i.e., they are not displayed in a red
background) and that in the evolution invalid operations, such as \texttt{1/0} or
\texttt{sqrt(-1)}, will not happen. The program will silently mask these invalid
results and provide instead some value, which will be almost always
unphysical.

\item The simulation may fail to find the maximum (or the minimum) value
allowed for $r$, especially if it happens for very large values of $r$ or
the potential oscillates badly. Try increasing $N$ and if this does not work
you may unselect \textbf{Limits} to remove the wrong display. This will not
hinder the numerical simulation of the motion, which however may be
difficult with forces not smooth enough.

\item You can always press the reset button   to recover the initial settings.

\end{itemize}



\end{document}